\documentstyle[aps,prd,psfig]{revtex}
\begin{document}
\title{Inertial forces in the Casimir effect with two moving plates}
\author{L. A. S. Machado and P. A. Maia Neto\thanks{e-mail: \tt pamn@if.ufrj.br}}
\address{Instituto de F\'{\i}sica, UFRJ, Caixa Postal 68528, 
21945-970 Rio de Janeiro, Brazil}
\date{\today}
\maketitle
\begin{abstract}
We combine linear response theory and dimensional regularization in order to derive the dynamical 
Casimir force in the low frequency regime. We consider two parallel plates moving along the 
normal direction in $D-$dimensional space. 
We assume the free-space values for the mass of each plate to be known, 
and obtain finite, separation-dependent mass corrections resulting from the 
combined effect of the two plates. 
The global mass correction is proportional to the static Casimir energy, in 
agreement with Einstein's law of equivalence between mass and energy for stressed rigid bodies. 
\end{abstract}

\begin{center}
PACS number(s): 12.20.Ds, 11.10.Kk, 42.50.Lc
\end{center}

\section{Introduction}

The Casimir force on moving plates has nontrivial properties, which have been 
studied in recent years. It provides for dissipation of the plate's mechanical
energy, an effect closely related to the emission of photon pairs~\cite{fulling}~\cite{general}. 
This coupling also gives rise to decoherence of the plate's motion~\cite{decoherence}. 
As far as the photon emission effect is concerned, it is interesting to assume that the 
plate oscillates with a frequency close to a resonance frequency of a cavity resonator, 
because the effect is greatly enhanced in this case~\cite{resonators}. 
However, to achieve such unusually high mechanical frequencies is a challenge not yet 
undertaken. More generally, dissipative effects involve field modes with frequencies  of the order
of the mechanical frequency, and are very small when the latter is well below the lowest cavity
frequency. 

On the other hand, dispersive effects result from the contribution of the entire field spectrum, and 
hence are less sensitive to the variation of the mechanical frequency. In this 
paper, we consider the Casimir forces on two parallel moving plates, in the 
nonrelativistic  and long-wavelength approximations. 
Photon production rates for this geometry have been derived in this limiting case~\cite{mundarain}. 
Here we focus on the dispersive force, and take an additional approximation:
we assume the motion to be slow in the time 
scale associated to the time-of-flight of light between the plates. 
In this limit, the main nontrivial corrections to the static Casimir force are 
inertial forces, i.e., contributions proportional to the accelerations of the 
two plates. This effect was analyzed in detail by Jaekel and Reynaud~\cite{jr1} 
in the case of a one-dimensional 
space. Here we consider $D$ spatial dimensions, and use dimensional regularization, as 
in the formalism employed by  Ambjorn and Wolfram for the static Casimir effect~\cite{ambjorn}.

The Casimir force and photon emission rates 
for a moving dielectric 
semi-infinite half-space (single interface)
in $D$ dimensions were also computed~\cite{eberlein}.
Generalizing the result of a previous work~\cite{inertia-one-plate}, 
Ref.~\cite{eberlein} obtained cut--off dependent mass corrections for the
moving interface, which renormalize the bare mass.  The corrections and  
the bare mass are both infinite (in the no cut-off limit), and 
when considered separately, unaccessible to 
measurements, only the total 
mass possessing a physical meaning. 
The  mass corrections we shall derive here have a different interpretation.
Rather than mass renormalization, they represent 
a  measurable physical effect, not present in the case of a single plane interface (or plate)
considered in~\cite{eberlein}\cite{inertia-one-plate}.
The corrections are finite, and depend on the distance between the plates. As required for consistency,
they vanish at the limit of large distance, leaving the place for the unperturbed free-space masses
of each plate. The free-space mass  eventually also incorporates the vacuum radiation
pressure coupling through a renormalization of a bare mass, in the line of 
Refs.~\cite{eberlein} and \cite{inertia-one-plate}. Here we skip this important problem to focus on the 
effect brought about by the presence of the second plate, and start from the experimentally known free-space
masses~\cite{footnote}.
The mass of a single plate may also be modified when its surface is  corrugated.
Mass corrections for one and two parallel corrugated plates were computed in the case of lateral motion~\cite{kardar}.   

This paper is organized as follows. In section 2, we derive most of the formal results (leaving 
technical aspects to three appendices), and some physical implications are discussed in section 3. 
Additional comments and conclusion are presented in section 4.

\section{Inertial Forces}

We consider a massless scalar field $\phi$ in $D-$dimensional space. The field obeys Dirichlet 
boundary conditions at the moving plates $x_1=\delta\!q_1(t)$ and $x_1=a+\delta\!q_2(t):$
\[
\phi(\delta\!q_1(t),x_2,...,x_D,t)=0=\phi(a+\delta\!q_2(t),x_2,...,x_D,t)
\]
Hence $\delta\!q_1(t)$ and $\delta\!q_2(t)$ represent the time-dependent deviations of the 
positions of the plates from their unperturbed values, which correspond to a $D$-dimensional 
static cavity of length $a.$ 

The vacuum radiation pressure force on plate $\alpha$ ($\alpha=1,2$ are the labels for the two boundaries)
is written as ${\hat f}^{(0)}_{\alpha}+\delta\!{\hat f}_{\alpha}(t),$
where ${\hat f}^{(0)}_{\alpha}$ is the static Casimir force operator 
for the unperturbed cavity length $a.$ 
We use hats to distinguish between the force operators and their average values.
For instance, the static {\it average} forces satisfy 
$f^{(0)}_{2}\doteq\langle {\hat f}^{(0)}_{2} \rangle = -f^{(0)}_{1},$
but no such simple relation holds for the force operators themselves.
We compute $\delta\!f_{\alpha}(t)$ up to first order in 
$\delta\!q_1(t)$ and $\delta\!q_2(t)$ and their time derivatives, 
from the 
fluctuations of the force in the unperturbed case, with the help of linear response theory.
In the frequency domain (we employ capital
letters to denote the Fourier transforms), we have 
\begin{equation}
\delta\!F_{\alpha}(\omega) = \sum_{\beta=1}^{2} \chi_{\alpha \beta} (\omega) \delta\!Q_{\beta}(\omega),\label{lin1}
\end{equation}
with the Fourier transform defined as:
\[
\delta\!F_{\alpha}(\omega)=\int d t \, \delta\!f_{\alpha}(t)\, e^{i\omega t}.
\]

The susceptibility $\chi_{\alpha \beta}(\omega)$ may be computed with the help of linear response
theory. In this approach, the response is derived from the fluctuations of the force in the 
static (unperturbed) case~\cite{jr_motional}. 
The imaginary part of the susceptibility, which accounts for dissipation, is given by
\begin{equation}
{\rm Im} \chi_{\alpha \beta}(\omega) = {1\over 2 \hbar} \left(C_{\alpha \beta}(\omega)-C_{\alpha \beta}(-\omega)\right),\label{rep1}
\end{equation}
where $C_{\alpha \beta}(\omega)$ is the Fourier transform of the correlation function of the 
force:
\[
c_{\alpha \beta}(t) = \langle  {\hat f}^{(0)}_{\alpha}(t) 
{\hat f}^{(0)}_{\beta}(0) \rangle -    f^{(0)}_{\alpha} f^{(0)}_{\beta}.
\]
The average is taken over the vacuum state, and for the unperturbed  configuration 
corresponding to the stationary cavity of length $a.$ 
The force is computed from the stress tensor  (we take $c=1$ in this section)
\begin{equation}
S_{i j}=-\partial_i \phi \,\partial_j\phi + {1\over 2} \delta_{ij} \left[\sum_{k=1}^D(\partial_k\phi)^2-
(\partial_t\phi)^2\right],
\end{equation}
with $i,j=1,...,D.$

When considering the fluctuations of ${\hat f}^{(0)}_{\alpha}$ we replace $\phi$ by the 
unperturbed field $\phi_0,$ which corresponds to the stationary configuration, and hence 
vanishes at $x_1=0$ and $x_1=a:$
\begin{equation}
\phi_0(x_1,...,x_D,t)=\sum_{l=1}^{\infty}\sum_{\{n_k\}} \sqrt{\hbar\over   \omega_{l,\{n_k\}}a L^{D-1}}
\sin\left({l \pi x_1\over a}\right) \exp(i {k}_{\{n_k\}}\cdot {x}_{\parallel}) a_{l,\{n_k\}} \label{normal}
\exp(-i \omega_{l,\{n_k\}} t) + {\rm H. c.},
\end{equation}
where ${x}_{\parallel}=(x_2,...,x_D),$ and 
 $\{n_k\}=\{n_2,...,n_D\}$ is a list of integer numbers (we have taken periodic  conditions 
at the boundaries of a cell of measure $L^{D-1}$ on the hyper-planes at $x_1=0$ and $x_1=a;$ as usual 
the limit $L\rightarrow \infty$ is assumed), and the corresponding sum is over all integer values. Each list
$\{n_k\}$ corresponds to a wavevector $k_{\{n_k\}}=2\pi (n_2,...,n_D)/L$ parallel to the hyper-planes.  
Hence a given field mode corresponds to an  integer $l$ and  a list $\{n_k\},$ and its 
associated frequency is $\omega_{l,\{n_k\}}=\sqrt{(l \pi/a)^2+k_{\{n_k\}}^2}.$

The force on the boundary at $x_1=0$ is given by
\begin{equation}
{\hat f}^{(0)}_1(t) = \int_{x_1=0}d^{D-1}x_{\parallel}\, S_{1,1}(0^+,x_{\parallel})  = -{1\over 2} 
 \int_{x_1=0}d^{D-1}x_{\parallel}\, \left(\partial_{x_1} \phi_0(0^+,x_{\parallel},t)\right)^2,\label{f1}
\end{equation}
whereas the force on the second boundary is
\begin{equation}
{\hat f}^{(0)}_2(t) =  {1\over 2} 
\int_{x_1=a}d^{D-1}x_{\parallel}\, \left(\partial_{x_1} \phi_0(a^-,x_{\parallel},t)\right)^2,\label{f2}
\end{equation}
the integrals taken over the quantization cell.
The spectra of force fluctuations are computed in Appendix A, yielding, with the help of (\ref{rep1}),
the imaginary parts of the four susceptibilities. They satisfy the symmetry relations
$ \chi_{11}(\omega)=\chi_{22}(\omega),$
$
\chi_{12}(\omega)=\chi_{21}(\omega).
$
We find 
\begin{eqnarray}
{\rm Im}\,\chi_{\alpha\beta}(\omega)&=&\frac{\pi^5\hbar L^{D-1}}{2a^{6}} \sum_{l_1=1}^{\infty}\label{imag}
\sum_{l_2=1}^{\infty}s_{\alpha\beta}(l_1,l_2)(l_1 l_2)^2
 \,\int\frac{d^{D-1}k_{\|}}{(2\pi)^{D-1}} 
{1\over \omega_{l_1}\omega_{l_2}}\\
 &\times&[ \delta(\omega-(\omega_{l_1}+\omega_{l_2}))-
\delta(\omega+(\omega_{l_1}+\omega_{l_2}))],
\end{eqnarray}
where $s_{11}(l_1,l_2)=1,$ $s_{12}(l_1,l_2)=-(-1)^{l_1+l_2},$ and 
\begin{equation}
\omega_{l}=\sqrt{(l\pi/a)^{2}+k_{\|}^{2} }.\label{omega}
\end{equation}
According to Eq.~(\ref{imag}),  
for a given mechanical frequency $\omega,$ dissipation comes from contributions of cavity modes
satisfying the resonance condition associated with the emission of photon pairs: 
$\omega_{l_1}+\omega_{l_2}=\omega.$ In fact, the mechanical power dissipated by the force 
equals the total radiated power, as expected by energy conservation~\cite{mundarain}. 

On the other hand, the real part of the susceptibility, which accounts for the dispersive 
component of the force, is computed from the dispersion relation~\cite{moyses}\cite{footnote2}
(${\rm P}\int$ denotes Cauchy's principal value)
\begin{equation}
{\rm Re}\,\chi_{\alpha\beta}(\omega) = {1\over \pi} {\rm P}\int d\omega' 
{{\rm Im}\,\chi_{\alpha\beta}(\omega')\over \omega'-\omega},\label{dispertion}
\end{equation}
and hence originates from contributions of the entire field spectrum.
Usually, mechanical frequencies are much smaller than the resonance frequencies of the cavity:
$\omega \ll \pi/L.$ In this limit, the motion is slow in the time scale corresponding to
the time-of-flight of the
photon between the plates. In this 
quasi static limit, the susceptibility may be replaced by 
the first terms of the expansion in powers of $\omega$ (we follow the notation of 
Ref.~\cite{jr1}):
\begin{equation}
\chi_{\alpha\beta}(\omega) = -\kappa_{\alpha\beta} + i\lambda_{\alpha\beta}\omega + \mu_{\alpha\beta}\omega^2 +{\cal O}(\omega^3) \label{quasistatic}
\end{equation}
The meaning of the coefficients in Eq.~(\ref{quasistatic}) is best understood in the time domain:
\begin{equation}\label{tempo}
\delta f_{\alpha}(t)= -\sum_{\beta=1}^{2} \left[\kappa_{\alpha\beta}\delta\!q_{\beta}(t) +
 \lambda_{\alpha\beta}\delta\!\dot q_{\beta}(t) + \mu_{\alpha\beta}\delta\!\ddot q_{\beta}(t)+...\right] .
\end{equation}

The coefficients $\lambda_{\alpha\beta}$ represent the viscous dissipative force in 
vacuum, in the low-frequency limit. 
They vanish for any value of $D,$
because there are no cavity modes available at low frequencies to contribute 
to ${\rm Im}\,\chi_{\alpha\beta}(\omega)$ as given 
by the r.-h.-s. of  Eq.~(\ref{imag}).

In contrast to the dissipative component of the force, the dispersive component 
is the combined  effect of all cavity modes even at such low mechanical frequencies.
It corresponds to the coefficients  $\kappa_{\alpha\beta}$ and $\mu_{\alpha\beta}$
in (\ref{quasistatic}) and (\ref{tempo}), which may be calculated from the low frequency expansion 
of (\ref{dispertion}). 
The coefficients $\kappa_{\alpha\beta}=-\chi_{\alpha\beta}(0)$ provide the  linear correction to the static Casimir 
force, when the distance between the plates is changed from $a$ to its 
instantaneous value $a-\delta q_1(t)+\delta q_2(t).$
Hence they do not contain any new nontrivial information, and are given by
\[
\kappa_{11}=-\kappa_{12}={\partial f^{(0)}_1\over \partial a}={\partial^2 E_0\over \partial a^2}
\]
where 
\begin{equation}
E_0 = -{\Gamma\left((D+1)/2\right)\,\zeta(D+1)\over (4\pi)^{(D+1)/2}}\, {\hbar L^{D-1}\over a^{D}}\label{E0}
\end{equation} 
is the (static) Casimir energy  for a cavity length $a$~\cite{ambjorn} ($\Gamma$ and 
$\zeta$ denote the gamma and Riemann zeta functions~\cite{abramowitz}).
Note that in the static case, only the 
relative position between the plates matters, so that $\chi_{11}(0)=-\chi_{12}(0).$ However, as we
show below, this relation does not hold for arbitrary values of frequency. 

Finally, $\mu_{\alpha\beta}$ are the coefficients of the inertial forces coupling the dynamics of the 
two plates. 
Replacing Eq.~(\ref{imag}) into (\ref{dispertion}), we find 
\begin{equation}
\mu_{\alpha\beta}=
\pi^{4} {\hbar L^{D-1}\over a^{6}} \sum_{l_1,l_2}
s_{\alpha\beta}(l_1,l_2)(l_1 l_2)^{2}\int {d^{D-1}k_{\|}\over (2\pi)^{D-1}} \frac{ 1 }{\omega_{l_1}\omega_{l_2}(\omega_{l_1}+\omega_{l_2})^3}.\label{mu11}
\end{equation}
We calculate the r.-h.-s. of (\ref{mu11}) in Appendix B, with the help of
dimensional regularization. We find
\begin{equation}
\mu_{11}=-{1\over 6 \pi^2}{\pi^{D/2}\over 2^{D}}\left[ {(D-3)(D-1)\over D}\Gamma\left(2-{D\over 2}\right)
\zeta(2-D) + \pi^{3/2-D} (D+1)\Gamma\left({D+1\over 2}\right)\zeta(D+1)
\right] \, {\hbar L^{D-1}\over a^D},\label{res11C}
\end{equation}
and
\begin{equation}
\mu_{12}=-{1\over 6 \pi^2}{\pi^{D/2}\over 2^{D}}\left[- {(D-3)(D-1)\over D}\Gamma\left(2-{D\over 2}\right)
\zeta(2-D) + {\pi^{3/2-D}\over 2} (D+1)\Gamma\left({D+1\over 2}\right)\zeta(D+1)
\right] \, {\hbar L^{D-1}\over a^D}.\label{res12C}
\end{equation}
Despite of the poles of the Gamma function, the first term in r.-h.-s. of (\ref{res11C}) and 
(\ref{res12C}) has a finite limit at positive even values of $D$ larger than $2,$ because of the 
zeros of the zeta function. Those values may be obtained more easily from the 
representation given by (\ref{res11B}) in Appendix B. Moreover, this term has a removable 
singularity  at $D=1,$ this time due to the pole of the zeta function, but again the limit is finite.
The 
Casimir energy $E_0,$
as well as the two inertial coefficients 
have negative values for any $D > 0,$ and 
in Fig.~1, we plot their 
absolute values  per unit of 
`area' as  functions of $D,$ 
and taking $\hbar=1,$ $a=1.$
They have minima between $D=22.8$ ($\mu_{12}$) and $D=25.2$ ($E_0$), and then increase exponentially
for larger values of $D.$

\begin{figure}
\centering \leavevmode
\psfig{file=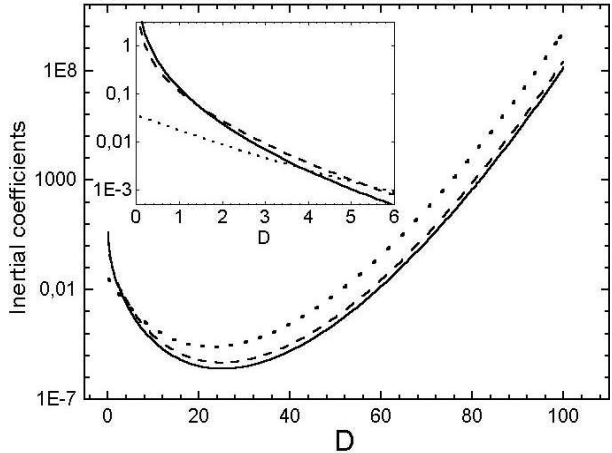,height=7.0cm,width=9.3cm}
\caption{Inertial coefficients and Casimir energy per unit area versus dimension. 
We take $\hbar=1$ and $a=1.$
Solid line: Casimir energy (absolute value), dashed line: 
$-\mu_{11},$ dotted line: $-\mu_{12}$}
\end{figure}

In Table~1, we show the explicit expressions for $D=1,$ $2$ and $3.$ 
The results for $D=1$ were already derived in Ref.~\cite{jr1}, 
as the linear and low-frequency limit of 
the exact result 
obtained by Fulling and Davies~\cite{fulling}, and also as 
the perfect-reflecting limit of the linear result for mirrors 
with frequency-dependent reflection coefficients. 
For $D=1$ and $D=3,$ we have also performed an independent calculation using an alternative 
regularization prescription, based on the introduction of a 
high-frequency cut-off, in order to check the results. 

\begin{table}
\caption{Inertial coefficients}
\begin{tabular}{cccc}
\LARGE
 &$D=1$ &$D=2$& $D=3$ \\ \tableline
 $\mu_{11}/(\hbar L^{D-1}/a^D)$& $-(3+\pi^2)/(36 \pi)$ &   $-(1+6\zeta(3))/(96 \pi)$ &      $-\pi^2/1080 $  \\                     
$\mu_{12}/(\hbar L^{D-1}/a^D)$&  $-(6+\pi^2)/(72\pi)$ &  $-(-1+3\zeta(3))/(96 \pi) $    &  $-\pi^2/2160$  \\
\end{tabular}
\end{table}

Since $\mu_{11}\neq -\mu_{12},$ the dynamical Casimir forces are not functions of the 
relative motion only, and the effect takes place also in the case of rigid motion of the 
cavity. This fact leads to a global mass correction, which is directly connected to the 
Casimir energy $E_0$ as we discuss in the next section.

\section{Global inertial correction and its connection with the Casimir energy.}

The inertial forces given by Eq.~(\ref{tempo}) leads to coupled equations of motion for the
two plates: 
\begin{equation}\label{eq_motion}
m_{\alpha} \delta\!\ddot q_{\alpha}(t)=
 -\sum_{\beta=1}^{2}  \mu_{\alpha\beta}\delta\!\ddot q_{\beta}(t)+ F_{\alpha},
\end{equation}
where $F_{\alpha}$ represents the sum of all forces acting on plate $\alpha,$
excluding the inertial force itself. 
The coefficients $\mu_{\alpha\beta}$ go to zero as $a^{-D}$ 
when the plates are set apart, hence
$m_{\alpha}$ is the free-space mass of plate $\alpha.$
Remarkably, the inertial force on a given plate also depends on the 
acceleration of the other plate. 
Even in the absence of external forces, in general
these equations cannot be reduced to uncoupled equations for center of mass and 
relative motion, unless $m_1=m_2=m,$ and in this case the 
effective mass for relative motion is
$(m+\mu_{11}-\mu_{12})/2.$

In the case of global, rigid motion of the `cavity' system, $\delta\! q_1= \delta\! q_2\doteq \delta\! q,$
we find,
by adding the equations of motion for the two plates,     
$
M \delta\! q'' = F_{\rm ext},
$
where $F_{\rm ext}$ is the sum of all external forces (the static 
Casimir components cancel, as well as the external stresses 
which enforce the rigidity by compensating the Casimir attraction), 
and 
\begin{equation}
M=m_1+m_2+2(\mu_{11}+\mu_{12}).\label{M}
\end{equation}
Note that $\mu_{11}+\mu_{12}$ is not a mass correction of a given plate taken 
separately, since $\mu_{12}$ is associated to a coupling between the plates as shown by
Eq.~(\ref{eq_motion}). However, $\Delta M=2(\mu_{11}+\mu_{12})$ is in fact a global 
inertial mass correction introduced by the Casimir effect, 
and from Eqs.~(\ref{E0}), (\ref{res11C}) and (\ref{res12C}) we find 
(now re-introducing the speed of light $c$)
\begin{equation}\label{mass-energy}
\Delta M= (D+1) {E_0\over c^2}.
\end{equation}
This relation between the inertial mass correction and the 
Casimir energy is in agreement with the  law of inertia of energy for a stressed rigid body.   
In fact, for a plane cavity (length $a$) moving along the normal direction, 
the general mass-energy relation reads~\cite{jr1}~\cite{einstein}
\begin{equation}\label{equivalence}
\Delta M = ( E +  a F )/c^2,
\end{equation}
where $E$ is the 
energy of the system contained within the cavity, and $ F $ the corresponding force, with 
$F>0$ denoting outward force~\cite{pauli}. Here $E=E_0$ is the unperturbed Casimir energy, and 
$F=f_2^{(0)}=-f_1^{(0)}$ is the static Casimir force.
From the power law as given by Eq.~(\ref{E0}) we find 
$$f_2^{(0)}=-{d E_0\over da}=D E_0/a,$$ which combined with (\ref{equivalence}) verifies Eq.~(\ref{mass-energy}). 

In view of  recent measurements of the Casimir force~\cite{mohideem}, we evaluate the 
order of magnitude of the inertial correction, taking $D=3$ and assuming that 
(\ref{mass-energy}) also holds for the electromagnetic case. 
Then the global mass correction per unit area is 
$${\Delta M\over L^2} = -{\pi^2 \hbar\over 180 c a^3}.$$
For a plate 
separation  $a=100 {\rm nm},$ we have $\Delta M/L^2=-2.0 \times 10^{-24} {\rm g/cm^2},$
a very small correction.

\section{Conclusion}

We have calculated the dynamical Casimir force for two moving plates in a D-dimensional space. 
We have employed linear response theory, which allows us to compute the dynamical force
from the spectrum of fluctuations  of the force for the static configuration. In this approach, 
the motion is taken to be a small perturbation, and terms of second or higher order in the displacements
of the plates (and their time derivatives) are neglected. 
Moreover, we have taken the quasi static approximation, expanding the linear susceptibility in powers of frequency.

In this limit, the dominant dynamical contributions are inertial forces. They have been derived 
from dispersion relations, 
and with the help of dimensional regularization. 
The inertial force on a given plate has two contributions: one proportional to its own
acceleration (coefficient $-\mu_{11}$), and one proportional to the acceleration of the companion plate
(coefficient  $-\mu_{12}$). For $D=3,$ we have $\mu_{11}=2 \mu_{12},$ but for larger dimensions 
the cross-acceleration force dominates over the self-acceleration force, as shown in Fig.~1. 
The two effects jointly contribute to the global mass correction, which in its turn is 
related to the Casimir energy as predicted by the law of inertia of energy for 
stressed rigid bodies.

\acknowledgments

We are grateful to C. Farina,
A. Lambrecht and S. Reynaud for 
discussions, and to CNPq, PRONEX and FAPERJ for partial financial 
support.

\appendix

\section{Dissipative component of the force}

In this appendix, we compute 
the spectrum of fluctuations of the force, $C_{\alpha\beta}(\omega),$ 
which is directly connected by linear response theory 
to ${\rm Im}\chi_{\alpha\beta}(\omega),$ as shown by Eq.~(\ref{rep1}).

According to Eq.~(\ref{f1}), the force operators are quadratic in the 
field operators. Hence, the force correlation function may be computed from  
two photon matrix elements as follows~\cite{barton}:
\begin{equation}
c_{\alpha\beta}(t) = {1\over 2} \sum_{\epsilon,\epsilon'} \langle 0| {\hat f}^{(0)}_{\alpha}(t)|\epsilon,\epsilon'\rangle
 \langle \epsilon,\epsilon'| {\hat f}^{(0)}_{\beta}(0)| 0 \rangle,\label{c1}
\end{equation}
where $|\epsilon,\epsilon'\rangle$ is a two-photon state, the label $\epsilon={l,n_2,...,n_D}$
representing a given field mode. The time dependence in Eq.~(\ref{c1}) is of the form 
$$\langle 0| {\hat f}^{(0)}_{\alpha}(t)|\epsilon,\epsilon'\rangle=
\exp[-i(\omega_{\epsilon}+\omega_{\epsilon'})t]\langle 0| {\hat f}^{(0)}_{\alpha}(0)|\epsilon,\epsilon'\rangle$$ 
for only annihilation operators contribute in this matrix element. 
Thus, the spectrum of fluctuations is given by 
\begin{equation}
C_{\alpha\beta}(\omega) = \pi\sum_{\epsilon,\epsilon'}
\langle 0| {\hat f}^{(0)}_{\alpha}(0)|\epsilon,\epsilon'\rangle
 \langle \epsilon,\epsilon'| {\hat f}^{(0)}_{\beta}(0)|0 \rangle
\delta(\omega-\omega_{\epsilon}-\omega_{\epsilon'}).\label{spec1}
\end{equation}
The  matrix elements are computed from the normal mode expansion as given by Eq.~(\ref{normal}):
\begin{equation}
\langle 0| {\hat f}^{(0)}_1(0)|\epsilon,\epsilon'\rangle=\pi^2 {\hbar\over a^3} {l l'\over \sqrt{\omega_{\epsilon}\omega_{\epsilon'}}}
\delta_{n_2,-n'_2}...\delta_{n_D,-n'_D}, \label{matrix1}
\end{equation}
\begin{equation}
\langle 0| {\hat f}^{(0)}_2(0)|\epsilon,\epsilon'\rangle=-\pi^2 {\hbar\over a^3} {l l'\over \sqrt{\omega_{\epsilon}\omega_{\epsilon'}}}
(-1)^{l+l'}\delta_{n_2,-n'_2}...\delta_{n_D,-n'_D}.\label{matrix2}
\end{equation}
We combine 
Eqs.~(\ref{rep1}),  (\ref{spec1}), (\ref{matrix1}) and 
replace 
\[
\sum_{\{n_k\}}=\left({L\over 2\pi}\right)^{D-1} \int d^{D-1}k.
\]
to derive the result for ${\rm Im}\chi_{11}(\omega)$ 
as given by (\ref{imag}). The result for 
${\rm Im}\chi_{12}(\omega)$ is very similar, except for the 
additional factor $-(-1)^{l+l'}$ present in (\ref{matrix2}) and not in 
(\ref{matrix1}).

\section{Derivation of the mass corrections }

In this appendix, we compute the 
series and integral in the r.-h.-s. of 
(\ref{mu11}), which 
give the results for the
mass coefficients $\mu_{\alpha\beta}.$
Since the integrand 
depends on ${\bf k}_{\|}$ only through its modulus,
we may replace the 
multiple integral by an integral over $k_{\|}=|{\bf k}_{\|}|,$ with~\cite{ambjorn}
\[
d^{D-1}k_{\|}={2 \pi^{(D-1)/2}\over \Gamma({(D-1)\over 2})} k_{\|}^{D-2} dk_{\|}.
\]
The terms with $l_1=l_2$ are simpler, and may be solved directly in terms 
of the Euler Beta function~\cite{gradshtejn}:
\begin{equation}
\mu_{11}[l_1=l_2]={\pi^{(D-3)/2}\over 2^{D+2}} {B((D-1)/2,-D/2+3)\over \Gamma((D-1)/2)} 
\left(\sum_{l=1}^{\infty} l^{D-2} \right)\label{muequal}
{\hbar L^{D-1}\over a^D},
\end{equation}
and $\mu_{12}[l_1=l_2]=-\mu_{11}[l_1=l_2].$
The series in the r.-h.-s. of (\ref{muequal}) converges when $D < 1$ to the value 
$\zeta(2-D);$ outside this range the regularized result is obtained by analytic 
continuation in $D,$ with the help of the well-known analytic extension of the 
zeta function. 
The terms with $l_1\neq l_2$ are computed in the following way: we first replace 
into (\ref{mu11}) the result 
\[
{1\over (\omega_{l_1}+\omega_{l_2})^3}={\omega_{l_1}^3-\omega_{l_2}^3-3 \omega_{l_1}^2\omega_{l_2}
+3\omega_{l_1}\omega_{l_2}^2\over \left[({\pi l_1\over a})^2- ({\pi l_2\over a})^2\right]^3},
\]
which follows from (\ref{omega}), and then
the resulting integrals may also be computed in terms of the Beta function:
\begin{equation}
\mu_{\alpha\beta}[l_1\neq l_2]=
{\pi^{(D-3)/2}\over 2^{D-2}}{1\over \Gamma((D-1)/2)}\times\sum_{l_1=1}^{\infty}\sum^*_{l_2}\Biggl\{(l_1^2-l_2^2)^{-3}
s_{\alpha\beta}(l_1,l_2)
\label{munot}
\end{equation}
\[
\times\left[ B\left({D-1\over 2},1-{D\over 2}\right) l_1^4 l_2^D + 
B\left({D+1\over 2},-{D\over 2}\right) l_1^2 l_2^{D+2}
- 3B\left({D-1\over 2},-{D\over 2}\right) l_1^{D+2} l_2^2 \right] \Biggr\}
{\hbar L^{D-1}\over a^D},
\]
where $\sum^*$ runs over all positive values of $l_2,$ except $l_2=l_1.$
In Appendix C, we regularize
the sums in Eq.~(\ref{munot}) once more with the help
of the analytic extension of the zeta function. The complete result reads
\begin{equation}
\mu_{11}=-{1\over 24 \pi^2}{\pi^{D/2}\Gamma(-D/2)\over 2^{D}}\left[ (D-3)(D-2)(D-1)\zeta(2-D) + 4\pi^2 (D+1)\zeta(-D)
\right] \, {\hbar L^{D-1}\over a^D},\label{res11A}
\end{equation}
and 
\begin{equation}
\mu_{12}=-{1\over 24 \pi^2}{\pi^{D/2}\Gamma(-D/2)\over 2^{D}}\left[-(D-3)(D-2)(D-1)\zeta(2-D) + 2\pi^2 (D+1)\zeta(-D)
\right] \, {\hbar L^{D-1}\over a^D}.\label{res12A}
\end{equation}
Due to the poles of the Gamma function, these results are at first-sight 
ill-defined for even positive values of $D.$ However, these poles are compensated by the zeros of the 
expression within brackets, and more useful representations may be derived with the help of 
the result~\cite{ambjorn}
\begin{equation}
\Gamma\left({s\over 2}\right) \zeta(s) = \pi^{s-1/2} \Gamma\left({1-s\over 2} \right) \zeta(1-s).\label{reflection1}        
\end{equation}
We employ (\ref{reflection1}) for the second term within brackets in (\ref{res11A}), whereas for
the first term we  employ the recurrence relation for the gamma function and then 
use (\ref{reflection1}) with $s=2-D.$ We find
\begin{equation}
\mu_{11}=-{1\over 12}{\pi^{-(D+1)/2}\over 2^{D}}\left[ -{(D-3)(D-2)(D-1)\over D}
\Gamma\left({D-1\over 2}\right)\zeta(D-1) + 2 (D+1)\Gamma\left({D+1\over 2}\right)
\zeta(D+1)
\right] \, {\hbar L^{D-1}\over a^D},\label{res11B}
\end{equation}
and a similar representation for $\mu_{12}.$
However, this representation is still undefined for $D=2,$ because of the pole of the zeta function. 
Hence a third representation is necessary to compute this particular case. Rather than  using 
(\ref{reflection1}) for the first term, we employ the recurrence relation twice to derive, again
from Eq.~(\ref{res11A}), the representation given by (\ref{res11C}).

\section{Series for the evaluation of $\mu_{11}$ and $\mu_{12}$}

In this appendix, we compute the series appearing in Eq.~(\ref{munot}). Our starting point
is the result
\begin{equation}
\sum_{l=-\infty}^{\infty} {l^2\over (l^2 - x^2)^3}=
{\pi\over 8 x^3} \left[{\pi x\over \sin^2(\pi x)} +
{1\over \tan(\pi x)}\left(1-{2\pi^2 x^2\over \sin^2(\pi x)}\right)\right]. 
           \label{der3}
\end{equation}
We evaluate the Laurent series expansions of the r.-h.-s. of (\ref{der3}) in powers of
$x-n,$ with $n$ integer, and from the lowest order terms derive the result
\begin{equation}
\sum^*_{l_2=1} {l_2^2\over (l_1^2-l_2^2)^3}= {1\over 32 l_1^4} - {\pi^2\over 48 l_1^2},\label{sum1}
\end{equation}
where the star denotes that the sum is taken from 1 to infinity, the value $l_2 = l_1$ excluded.
In a similar way we derive
\begin{equation}
\sum^*_{l_2=1} {l_2^4\over (l_1^2-l_2^2)^3}= -{1\over 32 l_1^4} - {5\pi^2\over 48}.
\end{equation}

For the derivation of $\mu_{12},$ we need 
\begin{equation}
\sum^*_{l_2=1} {(-1)^{l_2} l_2^2\over (l_1^2-l_2^2)^3}=
2\sum^*_{l_2=2,4,..} {l_2^2\over (l_1^2-l_2^2)^3}-\sum^*_{l_2=1} {l_2^2\over (l_1^2-l_2^2)^3}.
\end{equation}
For the first term we change the variable of sum to $k=l_2/2$ to find
\begin{equation}
\sum^*_{l_2=2,4,..} {l_2^2\over (l_1^2-l_2^2)^3}=
{1\over 16}\sum^*_{k=1} {k^2\over ((l_1/2)^2-k^2)^3}.\label{ult}
\end{equation}
For even values of $l_1$ this sum is given by the r.-h.-s. of (\ref{sum1}) (where we replace
$l_1$ by 
$l_1/2$), whereas for odd values the 
result may be obtained directly from  (\ref{der3}),  replacing $x$ by $l_1/2$ 
(in this case, the function in its
r.-h.-s. is finite at this value, and the sum in (\ref{ult}) runs over all positive integers). 
The final result reads
\begin{equation}
\sum^*_{l_2=1} {(-1)^{l_2} l_2^2\over (l_1^2-l_2^2)^3}=
(-1)^{l_1}\left( {1\over 32 l_1^4} + {\pi^2\over 96 l_1^2}\right). 
\end{equation}
Following the same approach, we compute
\begin{equation}
\sum^*_{l_2=1} {(-1)^{l_2} l_2^4\over (l_1^2-l_2^2)^3}=
(-1)^{l_1}\left( -{1\over 32 l_1^2} + {5\pi^2\over 96}\right). 
\end{equation}
After replacing these results in the r.-h.-s. of Eq.~(\ref{munot}), we still have  sums over 
$l_1$ to compute. The sums are divergent for the values of $D$ of physical interest, but 
regularized expressions are obtained by taking the analytic extension of the 
zeta function.


\begin{references}
\bibitem{footnote} This is loosely analogous to the standard approach in cavity quantum electrodynamics, where
one may start from the experimental free-space values for mass and charge (already renormalized)
when computing the corrections due to boundary effects.
\bibitem{footnote2} Note that this approach is not suitable for the problem with a single plate, as
discussed by P. A. Maia Neto and
S. Reynaud [Phys. Rev. A {\bf 47}, 1639 (1993)]. Here the proper high-frequency behavior essential to the 
validity of Eq.~(\ref{dispertion}) is provided by dimensional regularization.
\end{references}
\end{document}